\documentclass[twocolumn,preprintnumbers,amsmath,amssymb]{revtex4}

\usepackage[dvips]{graphicx}
\usepackage{bm}

\begin{document}

\title{Optimized stray-field-induced enhancement of the electron spin precession by buried Fe gates}

\author{L. Meier\footnote[1]{Also at: Solid State Physics Laboratory, ETH Zurich, 8093
Zurich, Switzerland}, G. Salis and N. Moll} \affiliation{IBM
Research, Zurich Research Laboratory, S\"aumerstrasse 4, 8803
R\"uschlikon, Switzerland}
\author{C. Ellenberger, I. Shorubalko, U. Wahlen and K. Ensslin}
\affiliation{Solid State Physics Laboratory, ETH Zurich, 8093
Zurich, Switzerland}

\author{E. Gini}
\affiliation{FIRST Center for Micro- and Nanosciences, ETH Zurich,
8093 Zurich, Switzerland}

\date{27. August 2007}

\begin{abstract}
The magnetic stray field from Fe gates is used to modify the spin
precession frequency of InGaAs/GaAs quantum-well electrons in an
external magnetic field. By using an etching process to position
the gates directly in the plane of the quantum well, the
stray-field influence on the spin precession increases
significantly compared with results from previous studies with
top-gated structures. In line with numerical simulations, the
stray-field-induced precession frequency increases as the gap
between the ferromagnetic gates is reduced. The inhomogeneous
stray field leads to additional spin dephasing.
\end{abstract}

\maketitle

The magnetic stray field from a confined ferromagnetic structure
is spatially strongly inhomogeneous. It decays on the length scale
of the magnetic object. Magnetic stray fields have many
applications ranging from data storage to future technologies such
as quantum computation~\cite{Loss1998} or
spintronics~\cite{Datta1990}. The spatial field distribution from
ferromagnetic structures has been characterized by magnetic-force
microscopes~\cite{Martin1987,WinklerJAP2006} or scanning Hall
probes~\cite{Chang1992}. The influence of stray fields on nearby
semiconductor spin-states has been investigated by
photoluminescence~\cite{Schomig2004,Sakuma2004}, spin-flip light
scattering~\cite{Sakuma2003},
cathodoluminescence~\cite{Kossut2001} and time-resolved Faraday
rotation~\cite{MeierAPL06, Meier2006} in semiconductor quantum
wells (QWs).

\begin{figure}
\includegraphics[width=80mm]{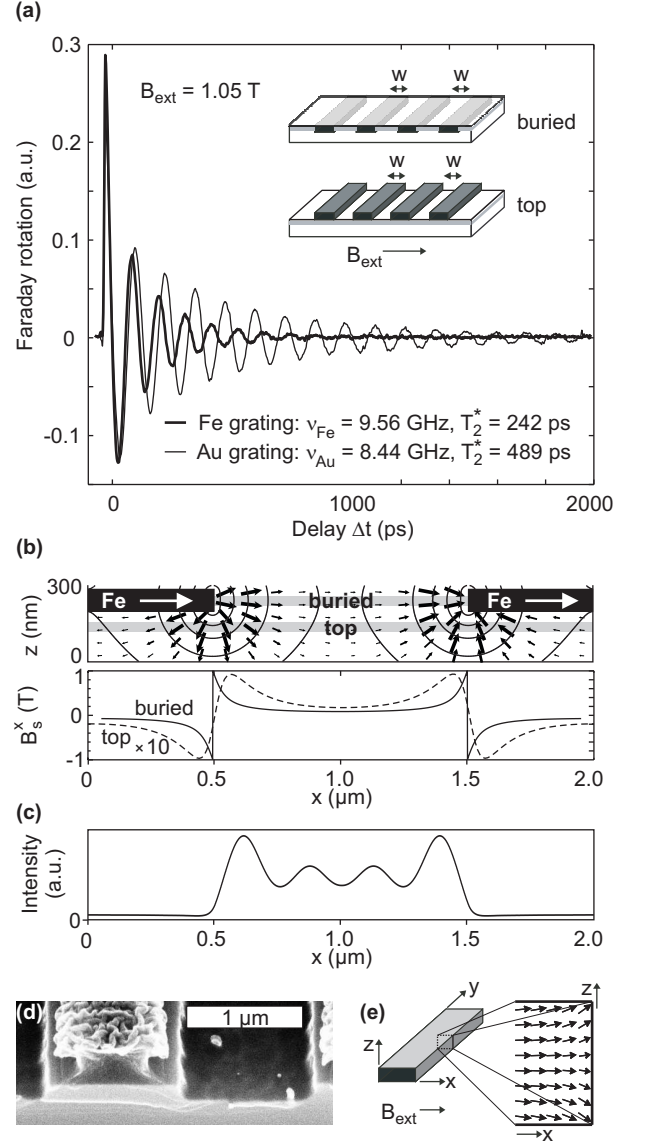}
\caption{\label{fig:fig1} (a) TRFR oscillations on a Fe (thick
line) and a Au (thin line) grating ($w \approx 1\,\mu$m). Inset:
sample structure with buried and top-gated Fe stripes. (b) Top:
Numerical simulation of the magnetic stray field $B_s$. Bottom:
Numerical simulation of the $x$-component of $B_s$ for buried and
top-gated sample structures. (c) Simulated illumination intensity
of the QW between two Fe stripes for the buried structure. (d)
Etched GaAs sample with double-layer PMMA mask. (e) Simulation of
the Fe stripe magnetization ($w=3\,\mu$m), in an external magnetic
field $B_\textrm{ext} = 500$~mT. Cut through the $x/z$ plane.}
\end{figure}

In previous experiments~\cite{MeierAPL06}, we reported that the
electron spin precession frequency $\nu$ of electrons in a QW
below an array of Fe stripes increases proportionally to the
component of the Fe magnetization perpendicular to the stripes and
in-plane of the QW. This enhancement of $\nu$ was less than
expected from a homogeneous average over the stray field in the
gap between two Fe stripes, and depended non-monotonically on the
geometry of the stripes. These discrepancies were
attributed~\cite{MeierAPL06,Meier2006} to near-field optical
effects and enhanced transmission close to the metallic stripes,
leading to the probing of electron spins not only in the gap
between two Fe stripes, but also below the stripes, where the
stray field points into the opposite direction of the external
magnetic field [Fig.~\ref{fig:fig1}{(b)}, situation `top'].

In this letter, we show that the stray-field-induced enhancement
of $\nu$ can be considerably increased if the Fe stripes are
brought into the plane of the QW by an etching process, thereby
eliminating the contributions of regions where the stray field
points against the external magnetic field. This stray-field
influence on $\nu$ is in good agreement with micromagnetic
stray-field simulations. We find a monotonous increase of this
enhancement when decreasing the width of the stripes and the gap
between two stripes, in line with theoretical expectations. The
spin decay-rate is increased by an amount that is proportional to
the stray-field-induced increase in precession frequency,
providing evidence of the occurrence of inhomogeneous broadening.

We use time-resolved Faraday rotation (TRFR)~\cite{Crooker1995} in
the Voigt geometry to measure the spin-precession frequency of
optically excited electron spins in a GaAs/InGaAs QW structure. A
first, circularly-polarized pump pulse (pulse duration 3~ps,
repetition rate 80~MHz at an average power of 400~$\mu$W, focus
diameter 15$~\mu$m) excites spin-polarized electrons into the
conduction band. The spin polarization initially points
perpendicularly to the QW plane, along the $z$-direction. After a
delay $\Delta t$, the spin polarization $S_z$ along $z$ is
monitored by measuring the angle $\theta_F \propto S_z$, by which
the polarization plane of a second, linearly polarized laser pulse
(pulse power 60~$\mu$W) is rotated. The signal obtained by
sweeping $\Delta t$ between 0 and 2~ns fits well to
$\theta_F(\Delta t) = \theta_0 \cos{(2\pi \nu \Delta t)}
\exp{(-t/T_2^\star)}$, where the exponential accounts for the
finite spin lifetime $T_2^\star$ and $\nu = g \mu_B
B_\textrm{tot}/h$ is proportional to the total magnetic field
$B_\textrm{tot}$ in the plane of the QW, with $g$ the electron
$g$-factor, $\mu_B$ the Bohr magneton and $h$ Planck's constant.
By measuring $\nu$, we can determine the total local magnetic
field $B_\textrm{tot} = B_\textrm{ext} + \langle B_s \rangle$
within the laser focus spot with a precision of $\approx 1$~MHz
(corresponding to $\approx 0.1$~mT). Here, $B_\textrm{ext}$ is an
externally applied magnetic field in the plane of the QW and
perpendicular to the stripe's long axis. The spatially
inhomogeneous stray field is averaged over the laser spot and
contributes with $\langle B_s \rangle$ to the total field. To
minimize additional effective magnetic fields from nuclear
polarization, we switch between left- and right-hand
circularly-polarized pump light at 50~kHz and measure at a
temperature of $T = 40$~K~\cite{MeierAPL06}.

Our sample is a 40-nm-wide GaAs/InGaAs QW (8.8\% In), with a bulk
Si-doping in the QW aimed at $5\times 10^{16}\,\textrm{cm}^{-3}$.
The QW is capped by 20~nm of GaAs, $\delta$-doped with Si to
compensate for surface states. In this case, the spin measurements
are not limited by the finite recombination time ($\approx
300$~ps) of optically excited electrons. Using electron-beam
lithography, arrays of stripes with a width $w$ and separated by a
distance $w$, $w \approx 1,2$ and $3\,\mu$m, are patterned into a
double-layered PMMA resist that is approximately 300~nm high. The
resulting gratings are 100 by 100~$\mu$m in size. The sample is
then etched with a reactive-ion etching (RIE) process with
7~standard cubic centimeters per minute (sccm) CH$_4$, 50~sccm
H$_2$, 5~sccm Ar, 3~sccm Cl$_2$ and an r.f. power of 135~W
(frequency 13.56 MHz) at $T = 95^\circ$C during twice 25~s,
interrupted by 5~min of N$_2$ flushing. The resulting 100-nm-deep
trenches [see Fig.~\ref{fig:fig1}{(d)}] are filled by the
evaporation of 10~nm Ti, 80~nm Fe and 10~nm Al. After a lift-off
process, the stripes are buried in the GaAs, centered about the QW
in $z$-direction, as schematically shown in the upper inset of
Fig.~\ref{fig:fig1}{(a)}.

A TRFR sweep with the laser focused on a grating ($w \approx
1\,\mu$m) with $B_\textrm{ext} = 1.05$~T is shown by the thick
line in Fig.~\ref{fig:fig1}{(a)}. Compared with a sample in which
the Fe has been replaced by Au (thin line), $\nu$ is enhanced by
$\Delta\nu = \nu_\textrm{Fe} - \nu_\textrm{Au} = 1.12$~GHz, which
is attributed to the magnetic stray field. Away from the magnetic
stripes, we determine $g=0.520$ and thus find $\langle B_s \rangle
= h\Delta\nu/g\mu_B = 154$~mT. This is about 8 times larger than
previous results on top-gated structures with the same stripe
geometry~\cite{MeierAPL06}. The spatially inhomogeneous $B_s$
leads to a distribution of precession frequencies in the spin
ensemble. The resulting dephasing rate $\tau^{-1}$ can be
approximated by $\tau2\pi\delta\nu = 1$, where $\delta\nu$ is an
effective spread of precession frequencies. It can be extracted
from the experiment by using $\tau^{-1} =
(T_\textrm{2,Fe}^\star)^{-1} -(T_\textrm{2,Au}^\star)^{-1}$, and
yields $\delta\nu \approx 334$~MHz, about three times smaller than
$\Delta\nu$.

A micromagnetic simulation of the magnetic stray field obtained
with OOMMF~\cite{oommf} is shown in Fig.~\ref{fig:fig1}{(b)}. In
the top panel, the geometrical orientation of the stray field
between two magnetized Fe stripes is shown for the structure with
QW between the buried stripes, and for the structure with the
stripes evaporated on top of the sample surface [see inset of
Fig.~\ref{fig:fig1}{(a)}]. The $x$-component of the stray field,
which our measurement geometry is mainly sensitive to, is shown in
the lower panel of Fig.~\ref{fig:fig1}{(b)}. The peak stray-field
in the samples with buried gates is approximately 10 times larger
than in the top-gated structures. A spatially homogeneous average
for the buried structure over $B_s^x(x)$ between the stripes
[$0.5\,\mu\textrm{m} < x \leq 1.5\,\mu\textrm{m}$ in
Fig.~\ref{fig:fig1}{(b)}] yields $\langle B_s \rangle = 107$~mT
for $w = 1\,\mu$m and $B_\textrm{ext}$~=~1~T.

\begin{figure}
\includegraphics[width=80mm]{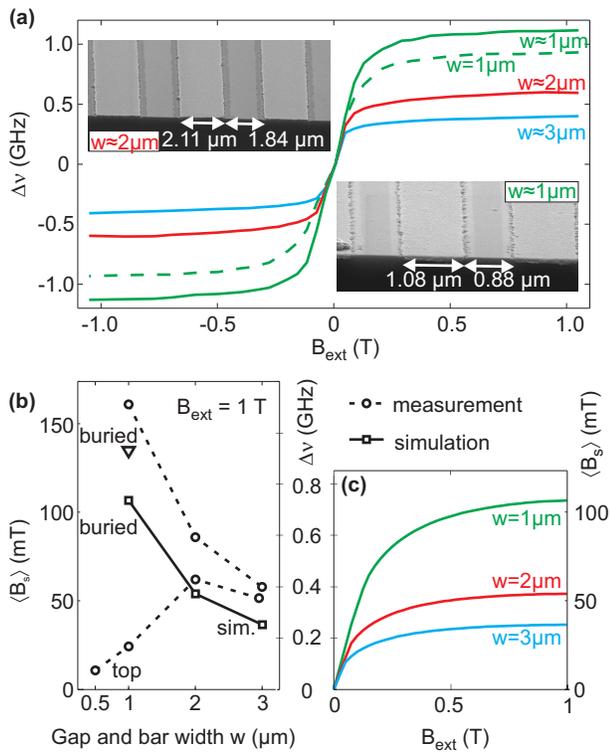}
\caption{\label{fig:fig2} (Color online) (a) Enhancement of the
electron spin precession frequency $\Delta\nu$ as a function of
$B_\textrm{ext}$ for different gap and stripe widths $w$ (solid
lines). Inset: Because of the etching process, the stripes are
slightly too wide and the gaps too narrow. The dashed line shows
$\Delta\nu$ for a sample with equal gap and stripe width
$w=1\,\mu$m. (b) $\langle B_s \rangle$ and $\Delta\nu$ for
different $w$ at $B_\textrm{ext} = 1$~T as measured (dashed lines)
for buried and top-gated and as simulated (solid line) for buried
structures. Triangle: Measured $\langle B_s \rangle$ for the
compensated $w=1\,\mu$m sample. (c) Simulated $\langle B_s
\rangle$ and $\Delta\nu$ as a function of $B_\textrm{ext}$ for
different $w$ (buried structures).}
\end{figure}

Figure~\ref{fig:fig2}{(a)} shows the measured $\Delta\nu$ as a
function of $B_\textrm{ext}$. At $B_\textrm{ext} = 0$, the Fe
stripes are magnetized along their long (easy) axis in the
$y$-direction, and no stray field in the $x$-direction is
expected. As $B_\textrm{ext}$ increases, the magnetization turns
towards $x$. For smaller stripe widths $w$, larger
$B_\textrm{ext}$ are needed to magnetize the Fe along $x$ because
of shape anisotropy. This is verified by MOKE
measurements~\cite{MeierAPL06}, and reflected in the shape of
$\Delta\nu$ vs.\ $B_\textrm{ext}$. This effect is also visible in
the simulation in Fig.~\ref{fig:fig2}{(c)}. Here, the
magnetization \boldsymbol{$m$} of the Fe stripes has been
calculated as a function of $B_\textrm{ext}$, and from
\boldsymbol{$m$}, we obtain the homogeneously averaged stray
field. After a roughly linear increase of $\langle B_s \rangle$
due to the rotation of \boldsymbol{$m$} from along $y$ into the
$x/z$ plane, $\langle B_s \rangle$ slowly increases. This increase
is due to \boldsymbol{$m$}, which, in the corners of the $x/z$
plane [see Fig.~\ref{fig:fig1}{(e)}], has components along $z$
that decrease as $B_\textrm{ext}$ is increased. If all magnetic
moments in the Fe stripes point along the $x$-direction, i.e.
$\boldsymbol{m} = m \boldsymbol{\hat{x}}$, the value of $\langle
B_s \rangle$ saturates.

The measured and calculated values of $\Delta\nu$ and $\langle B_s
\rangle$ at $B_\textrm{ext} = 1$~T are plotted in
Fig.~\ref{fig:fig2}{(b)} for varying gap and stripe widths $w$. In
the structure with buried stripes, the stray-field-induced
enhancement of $\nu$ decreases with $w$ in both measurement and
simulation. Because the magnetic stray field between two Fe gates
stems from the finite divergence of the magnetization at its
boundary, here mainly in the $y/z$ plane, the gate's extension in
the $x$-direction has only a minor influence on $B_s$. The average
$\langle B_s \rangle$ is predominated by the high values of
$B_s(x)$ for $x$ close to a stripe edge, which is independent of
the stripe width. With increasing $w$, $\langle B_s \rangle$
decreases roughly as $1/w$ because more and more of the gap region
sees a negligible stray-field strength. In our previous work with
top-gated structures~\cite{MeierAPL06}, we observed a
non-monotonous dependence of $\Delta\nu$ on $w$ [see the dashed
line `top' in Fig.~\ref{fig:fig2}{(b)}] owing to the probing of
negative stray fields in regions below the stripe, which were
illuminated because of effects due to enhanced optical
transmission.

While the qualitative behavior of the buried gates is similar in
experiment and simulation, the absolute value of $\langle B_s
\rangle$ is about $50\%$ higher in the experiment than in the
simulation. This is partly due to the fact that the RIE process
etches the mask also slightly laterally, leading to stripes that
are $\approx 100$~nm wider than the original mask, at the cost of
narrower gaps [see the SEM images in the insets of
Fig.~\ref{fig:fig2}{(a)}]. By using a compensated mask geometry
with narrower stripes, the resulting $\langle B_s \rangle$
decreases in the experiment, as shown by the dashed line in
Fig.~\ref{fig:fig2}{(a)} and the triangle in
Fig.~\ref{fig:fig2}{(b)} for a $w=1\,\mu$m grating.

Still, $\langle B_s \rangle$ exceeds the numerically expected
values by about $25\%$. This may be explained by a non-homogeneous
averaging of the stray field in the gap. Numerical simulations of
the light intensity distribution [finite-difference time-domain,
Fig.~\ref{fig:fig1}{(c)}] reveal an enhanced intensity close to a
stripe edge. Therefore, these regions with higher-than-average
stray fields contribute more to the spatial average and may lead
to an elevated $\langle B_s \rangle$. However, the simulation
predicts a sharp drop of the light intensity at the metal edge as
well, which leads to an underrepresentation of the peak
stray-field values (compared with a homogeneous average) and
neutralizes the former effect. We note that also the sensitivity
of TRFR angle on spin polarization may be spatially modulated
owing to near-field optical effects, leading to additional
weighting of the stray-field average. A detailed description of
this effect would exceed the scope of this paper.

Also the spin lifetimes exhibit a strong dependence on $w$. As
described previously, the additional decoherence rate $\tau^{-1}$
and thereby the stray-field magnitude can be estimated from
$T_2^\star$. We find $\delta\nu = 334, 191$ and $72$~MHz for the
$w\approx 1, 2$ and $3\,\mu$m gratings, respectively. Although the
total spread of the stray field is independent of $w$, the
effective spread $\delta\nu$ is diluted by increasing $w$ (because
more volume with negligible stray field is added), leading to a
decrease of $\delta\nu$, in line with $\Delta\nu$ determined from
the spin-precession frequency.

In conclusion, we have measured the effect of magnetized Fe gates
on the electron spin-precession frequency $\nu$. By etching the
gates down to the quantum well, the numerically expected average
stray field $\langle B_s \rangle$ between two stripes is increased
by a factor of $\approx 3$, compared with the situation with
top-gated Fe electrodes. In the experiment, $\nu$ is increased by
up to one order of magnitude for samples with buried gates,
because here near-field optical effects do not lead to the probing
of regions with stray fields pointing along the negative
$x$-direction, as it is the case for top-gated structures.


\begin{thebibliography}{13}
\expandafter\ifx\csname
natexlab\endcsname\relax\def\natexlab#1{#1}\fi
\expandafter\ifx\csname bibnamefont\endcsname\relax
  \def\bibnamefont#1{#1}\fi
\expandafter\ifx\csname bibfnamefont\endcsname\relax
  \def\bibfnamefont#1{#1}\fi
\expandafter\ifx\csname citenamefont\endcsname\relax
  \def\citenamefont#1{#1}\fi
\expandafter\ifx\csname url\endcsname\relax
  \def\url#1{\texttt{#1}}\fi
\expandafter\ifx\csname
urlprefix\endcsname\relax\def\urlprefix{URL }\fi
\providecommand{\bibinfo}[2]{#2}
\providecommand{\eprint}[2][]{\url{#2}}

\bibitem[{\citenamefont{Loss and DiVincenzo}(1998)}]{Loss1998}
\bibinfo{author}{\bibfnamefont{D.}~\bibnamefont{Loss}} \bibnamefont{and}
  \bibinfo{author}{\bibfnamefont{D.~P.} \bibnamefont{DiVincenzo}},
  \bibinfo{journal}{Phys. Rev. A} \textbf{\bibinfo{volume}{57}},
  \bibinfo{pages}{120} (\bibinfo{year}{1998}).

\bibitem[{\citenamefont{Datta and Das}(1990)}]{Datta1990}
\bibinfo{author}{\bibfnamefont{S.}~\bibnamefont{Datta}} \bibnamefont{and}
  \bibinfo{author}{\bibfnamefont{B.}~\bibnamefont{Das}},
  \bibinfo{journal}{Appl. Phys. Lett.} \textbf{\bibinfo{volume}{56}},
  \bibinfo{pages}{665} (\bibinfo{year}{1990}).

\bibitem[{\citenamefont{Martin and Wickramasinghe}(1987)}]{Martin1987}
\bibinfo{author}{\bibfnamefont{Y.}~\bibnamefont{Martin}} \bibnamefont{and}
  \bibinfo{author}{\bibfnamefont{H.~K.} \bibnamefont{Wickramasinghe}},
  \bibinfo{journal}{Appl. Phys. Lett.} \textbf{\bibinfo{volume}{50}},
  \bibinfo{pages}{1455} (\bibinfo{year}{1987}).

\bibitem[{\citenamefont{Winkler et~al.}(2006)\citenamefont{Winkler, Muhl,
  Menzel, Kozhuharova-Koseva, Hampel, Leonhardt, and Buchner}}]{WinklerJAP2006}
\bibinfo{author}{\bibfnamefont{A.}~\bibnamefont{Winkler}},
  \bibinfo{author}{\bibfnamefont{T.}~\bibnamefont{Muhl}},
  \bibinfo{author}{\bibfnamefont{S.}~\bibnamefont{Menzel}},
  \bibinfo{author}{\bibfnamefont{R.}~\bibnamefont{Kozhuharova-Koseva}},
  \bibinfo{author}{\bibfnamefont{S.}~\bibnamefont{Hampel}},
  \bibinfo{author}{\bibfnamefont{A.}~\bibnamefont{Leonhardt}},
  \bibnamefont{and} \bibinfo{author}{\bibfnamefont{B.}~\bibnamefont{Buchner}},
  \bibinfo{journal}{J. Appl. Phys.} \textbf{\bibinfo{volume}{99}},
  \bibinfo{eid}{104905} (\bibinfo{year}{2006}).

\bibitem[{\citenamefont{Chang et~al.}(1992)\citenamefont{Chang, Hallen,
  Harriott, Hess, Kao, Kwo, Miller, Wolfe, van~der Ziel, and
  Chang}}]{Chang1992}
\bibinfo{author}{\bibfnamefont{A.~M.} \bibnamefont{Chang}},
  \bibinfo{author}{\bibfnamefont{H.~D.} \bibnamefont{Hallen}},
  \bibinfo{author}{\bibfnamefont{L.}~\bibnamefont{Harriott}},
  \bibinfo{author}{\bibfnamefont{H.~F.} \bibnamefont{Hess}},
  \bibinfo{author}{\bibfnamefont{H.~L.} \bibnamefont{Kao}},
  \bibinfo{author}{\bibfnamefont{J.}~\bibnamefont{Kwo}},
  \bibinfo{author}{\bibfnamefont{R.~E.} \bibnamefont{Miller}},
  \bibinfo{author}{\bibfnamefont{R.}~\bibnamefont{Wolfe}},
  \bibinfo{author}{\bibfnamefont{J.}~\bibnamefont{van~der Ziel}},
  \bibnamefont{and} \bibinfo{author}{\bibfnamefont{T.~Y.} \bibnamefont{Chang}},
  \bibinfo{journal}{Appl. Phys. Lett.} \textbf{\bibinfo{volume}{61}},
  \bibinfo{pages}{1974} (\bibinfo{year}{1992}).

\bibitem[{\citenamefont{Sch\"{o}mig et~al.}(2004)\citenamefont{Sch\"{o}mig,
  Forchel, Halm, Bacher, Puls, and Henneberger}}]{Schomig2004}
\bibinfo{author}{\bibfnamefont{H.}~\bibnamefont{Sch\"{o}mig}},
  \bibinfo{author}{\bibfnamefont{A.}~\bibnamefont{Forchel}},
  \bibinfo{author}{\bibfnamefont{S.}~\bibnamefont{Halm}},
  \bibinfo{author}{\bibfnamefont{G.}~\bibnamefont{Bacher}},
  \bibinfo{author}{\bibfnamefont{J.}~\bibnamefont{Puls}}, \bibnamefont{and}
  \bibinfo{author}{\bibfnamefont{F.}~\bibnamefont{Henneberger}},
  \bibinfo{journal}{Appl. Phys. Lett.} \textbf{\bibinfo{volume}{84}},
  \bibinfo{pages}{2826} (\bibinfo{year}{2004}).

\bibitem[{\citenamefont{Sakuma et~al.}(2004)\citenamefont{Sakuma, Hykomi,
  Souma, Murayama, and Oka}}]{Sakuma2004}
\bibinfo{author}{\bibfnamefont{M.}~\bibnamefont{Sakuma}},
  \bibinfo{author}{\bibfnamefont{K.}~\bibnamefont{Hykomi}},
  \bibinfo{author}{\bibfnamefont{I.}~\bibnamefont{Souma}},
  \bibinfo{author}{\bibfnamefont{A.}~\bibnamefont{Murayama}}, \bibnamefont{and}
  \bibinfo{author}{\bibfnamefont{Y.}~\bibnamefont{Oka}},
  \bibinfo{journal}{Appl. Phys. Lett.} \textbf{\bibinfo{volume}{85}},
  \bibinfo{pages}{6203} (\bibinfo{year}{2004}).

\bibitem[{\citenamefont{Sakuma et~al.}(2003)\citenamefont{Sakuma, Hyomi, Souma,
  Murayama, and Oka}}]{Sakuma2003}
\bibinfo{author}{\bibfnamefont{M.}~\bibnamefont{Sakuma}},
  \bibinfo{author}{\bibfnamefont{K.}~\bibnamefont{Hyomi}},
  \bibinfo{author}{\bibfnamefont{I.}~\bibnamefont{Souma}},
  \bibinfo{author}{\bibfnamefont{A.}~\bibnamefont{Murayama}}, \bibnamefont{and}
  \bibinfo{author}{\bibfnamefont{Y.}~\bibnamefont{Oka}}, \bibinfo{journal}{J.
  Appl. Phys.} \textbf{\bibinfo{volume}{94}}, \bibinfo{pages}{6423}
  (\bibinfo{year}{2003}).

\bibitem[{\citenamefont{Kossut et~al.}(2001)\citenamefont{Kossut, Yamakawa,
  Nakamura, Cywi\'{n}ski, Fronc, Czeczott, Wr\'{o}bel, Kyrychenko, Wojtowicz,
  and Takeyama}}]{Kossut2001}
\bibinfo{author}{\bibfnamefont{J.}~\bibnamefont{Kossut}},
  \bibinfo{author}{\bibfnamefont{I.}~\bibnamefont{Yamakawa}},
  \bibinfo{author}{\bibfnamefont{A.}~\bibnamefont{Nakamura}},
  \bibinfo{author}{\bibfnamefont{G.}~\bibnamefont{Cywi\'{n}ski}},
  \bibinfo{author}{\bibfnamefont{K.}~\bibnamefont{Fronc}},
  \bibinfo{author}{\bibfnamefont{M.}~\bibnamefont{Czeczott}},
  \bibinfo{author}{\bibfnamefont{J.}~\bibnamefont{Wr\'{o}bel}},
  \bibinfo{author}{\bibfnamefont{F.}~\bibnamefont{Kyrychenko}},
  \bibinfo{author}{\bibfnamefont{T.}~\bibnamefont{Wojtowicz}},
  \bibnamefont{and} \bibinfo{author}{\bibfnamefont{S.}~\bibnamefont{Takeyama}},
  \bibinfo{journal}{Appl. Phys. Lett.} \textbf{\bibinfo{volume}{79}},
  \bibinfo{pages}{1789} (\bibinfo{year}{2001}).

\bibitem[{\citenamefont{Meier et~al.}(2006{\natexlab{a}})\citenamefont{Meier,
  Salis, Ellenberger, Ensslin, and Gini}}]{MeierAPL06}
\bibinfo{author}{\bibfnamefont{L.}~\bibnamefont{Meier}},
  \bibinfo{author}{\bibfnamefont{G.}~\bibnamefont{Salis}},
  \bibinfo{author}{\bibfnamefont{C.}~\bibnamefont{Ellenberger}},
  \bibinfo{author}{\bibfnamefont{K.}~\bibnamefont{Ensslin}}, \bibnamefont{and}
  \bibinfo{author}{\bibfnamefont{E.}~\bibnamefont{Gini}},
  \bibinfo{journal}{Appl. Phys. Lett.} \textbf{\bibinfo{volume}{88}},
  \bibinfo{eid}{172501} (\bibinfo{year}{2006}).

\bibitem[{\citenamefont{Meier et~al.}(2006{\natexlab{b}})\citenamefont{Meier,
  Salis, Ellenberger, Gini, and Ensslin}}]{Meier2006}
\bibinfo{author}{\bibfnamefont{L.}~\bibnamefont{Meier}},
  \bibinfo{author}{\bibfnamefont{G.}~\bibnamefont{Salis}},
  \bibinfo{author}{\bibfnamefont{C.}~\bibnamefont{Ellenberger}},
  \bibinfo{author}{\bibfnamefont{E.}~\bibnamefont{Gini}}, \bibnamefont{and}
  \bibinfo{author}{\bibfnamefont{K.}~\bibnamefont{Ensslin}},
  \bibinfo{journal}{Phys. Rev. B} \textbf{\bibinfo{volume}{74}},
  \bibinfo{pages}{245318} (\bibinfo{year}{2006}).

\bibitem[{\citenamefont{Crooker et~al.}(1995)\citenamefont{Crooker, Awschalom,
  and Samarth}}]{Crooker1995}
\bibinfo{author}{\bibfnamefont{S.~A.} \bibnamefont{Crooker}},
  \bibinfo{author}{\bibfnamefont{D.~D.} \bibnamefont{Awschalom}},
  \bibnamefont{and} \bibinfo{author}{\bibfnamefont{N.}~\bibnamefont{Samarth}},
  \bibinfo{journal}{IEEE J. Sel. Top. Quantum Electron.}
  \textbf{\bibinfo{volume}{1}}, \bibinfo{pages}{1082} (\bibinfo{year}{1995}).

\bibitem[{oom()}]{oommf}
\bibinfo{howpublished}{\url{http://math.nist.gov/oommf/}}.

\end{thebibliography}

\end{document}